# A Serious Game Design: Nudging Users' Memorability of Security Questions

**Nicholas Micallef**
Australian Centre for Cyber Security
School of Engineering and Information Technology
University of New South Wales
Canberra, Australia
Email: n.micallef@adfa.edu.au

**Nalin Asanka Gamagedara Arachchilage**
Australian Centre for Cyber Security
School of Engineering and Information Technology
University of New South Wales
Canberra, Australia
Email: nalin.asanka@adfa.edu.au

## Abstract

Security questions are one of the techniques used to recover passwords. The main limitation of security questions is that users find strong answers difficult to remember. This leads users to trade-off security for the convenience of an improved memorability. Previous research found that increased fun and enjoyment can lead to an enhanced memorability, which provides a better learning experience. Hence, we empirically investigate whether a serious game has the potential of improving the memorability of strong answers to security questions. For our serious game, we adapted the popular "*4 Pics 1 word*" mobile game because of its use of pictures and cues, which psychology research found to be important to help with memorability. Our findings indicate that the proposed serious game could potentially improve the memorability of answers to security questions. This potential improvement in memorability, could eventually help reduce the trade-off between usability and security in fall-back authentication.

# 1   Introduction

As internet users have to deal with an increasing number of online accounts, they are facing an enormous challenge to remember all passwords chosen for their accounts (Stavova et al. 2016). For example, if we just look at social networking sites, plenty of users have different accounts for Facebook, Twitter, Instagram and SnapChat. Since password managers have not been widely adopted (Alkaldi and Renauld 2016), resetting passwords is becoming a more frequent task (Florencio and Herley 2007; Stavova et al. 2016). To address this problem, various forms of fall-back authentication mechanisms have been deployed and evaluated with the most popular being security questions (Schechter and Reeder 2009) and email-based password recovery. Although email-based password recovery has been widely adopted by major companies (e.g. Google), they still have the limitation of being vulnerable to 'man in the middle' attacks, since these emails are not usually encrypted (Stavova et al. 2016).

Even security questions have several vulnerabilities. For instance, some answers to security questions can be easily accessible with a quick Google search (e.g. in Sarah Palin's 2008 email hack (Bridis 2008), the hacker merely used social engineering techniques to reset Palin's password using her birth-date, ZIP code and where she met her spouse). Also, since 2008, more of our personal information has become available online. Hence, it is becoming easier for attackers to retrieve this information (Golla and Dürmuth 2015), through observational attacks (i.e. the art of human hacking used to obtain sensitive information such as usernames, passwords, personal identification numbers through observing how victims behave both online and off-line), from social networking websites, such as LinkedIn or Facebook (Rabkin 2008). Besides observational attacks, security questions are also vulnerable to guessing attacks (Golla and Dürmuth 2015), in which, attackers try to access accounts by providing low entropy (i.e. the level of complexity) answers (e.g. favourite colour: blue). Thus, the ease of conducting observational and guessing attacks has increased the vulnerabilities of security questions (Just and Aspinall 2010) towards all these cyber-threats, which lead to severe consequences, such as monetary loss and embarrassment (Micallef and Just 2011).

A possible way to reduce the vulnerability of security questions towards cyber-attacks is to use system-generated answers (Micallef and Just 2011). However, the main barrier towards widespread adoption of these techniques is memorability (Just and Aspinall 2009), since users struggle to remember system-generated information to answer their security questions (Bonneau et al. 2015). Hence, the main challenge of the current implementation of security questions is that strong answers to security questions (i.e. high entropy), like those provided by system-generated answers (Micallef and Just 2011), are less prone to observational and guessing attacks, but at the same time are difficult for the user to remember the answers (Shay et al. 2012). Alternatively, weak answers to security questions (i.e. low entropy) are more prone to cyber-attacks (Bonneau et al. 2010; Denning et al. 2011), nevertheless they are easy for the user to remember the answers (Zviran and Haga 1990).

Since system-generated answers to security questions can limit the vulnerabilities to guessing and social engineering attacks (Shay et al. 2012), they seem to be the most promising solution to bridge the trade-off between usability and security in fall-back authentication (Micallef and Arachchilage 2017a, 2017b). However, system-generated answers to security questions need to be better presented to the user, by investigating techniques that could enhance memorability (Micallef and Arachchilage 2017a, 2017b). Previous work (Ho et al. 2009) found that an increase in fun and enjoyment leads to an enhanced memorability and consequently an improved learning experience. One could argue that an improved learning experience provides a high user satisfaction, which was empirically investigated by previous work on security behaviour (Arachchilage et al. 2016), that it could motivate users to change peoples' phishing threat avoidance behaviour. Hence, in this paper, we evaluate a serious game that nudges users' memorability of answers to security questions with the ultimate goal of bridging the trade-off between usability (i.e. memorability) and security during fall-back authentication.

We decided to implement a serious game as a mobile app and to primarily focus on the 18-35 age group because 95% of Australians in this age group own a mobile phone (Poushter 2016), due to its mobility nature. Hence, in our research, we adapted the popular picture-based "*4 Pics 1 Word*" mobile game (Google Play 2014). This game asks users to pick the word that relates the given pictures (e.g. for pictures in Figure 1a the relating word would be "Walk"). We selected this game because of its use of pictures and cues, in which, previous psychology research has found to be important to help with memorability (Paivio et al. 1968). Besides asking users to solve the standard game's challenges, we adapted this game, so that it also asks users to solve challenges based on the answers chosen to their security questions (security questions challenges) (Micallef and Arachchilage 2017a, 2017b).

Since the aim of this work is to understand whether or not the proposed serious game has the potential of improving the memorability of stronger answers to security questions (in this case system-generated answers), in our evaluation we assigned participants to one of two groups. Group 1 selects their own answers to security questions (i.e. control group). Group 2 uses answers based on a system-generated profile (i.e. experimental group) (similar to Micallef and Just (2011)). We also want to understand the short-term impact (in terms of workload and memorability) of using system-generated profiles, when memorizing answers to security questions, playing a game and then remembering answers to security questions. Hence, our research contributes to the field of fall-back authentication by answering the following research questions (RQs):

**RQ1:** Does the serious game have the potential of helping users remember stronger answers to security questions?

**RQ2:** What's the short-term impact of using a system-generated profile on memorizing answers to security questions, playing a game and then remembering answers to security questions?

We know to our cost that no-one has evaluated the design of a serious game to nudge users' memorability towards stronger answers to security questions based on system-generated profiles. In Section 2, we describe the background related to our research. In Section 3, we describe the game, security questions and system-generated profiles that we use in our user evaluation. Afterwards, we describe our user evaluation (Section 4), present the results of the evaluation (Section 5) and discuss how these results answer **RQ1** and **RQ2** (Section 6). Finally, we present the limitations of this research, future work (Section 7) and main conclusions that could be drawn from this work (Section 8).

## 2 Related Work

### 2.1 Security Questions and System-generated Data

Security questions are set-up at account creation. Then, at password recovery, users have to remember the answers that they provided when setting up the account. Recent studies, conducted using security questions data collected by Google (Bonneau et al. 2015), found that security questions are neither usable (i.e. low memorability) nor secure enough to recovery passwords. This means that new techniques need to be investigated to provide more secure and memorable security questions.

System-generated password schemes were evaluated to be more secure than user-defined passwords (Shay et al. 2012). However, they were evaluated to be not memorable (Al-Ameen et al. 2015), even when using natural-language words (Wright et al. 2012). For instance, Wright et al. (2012) evaluated the usability of three system-generated password schemes and found that these schemes did not have sufficient memorability rates. Also, Forget et al. (2008) evaluated a hybrid scheme which uses both user-selected and system-generated passwords by having a system which randomly adds characters to a user-chosen password to improve its' security. This scheme only achieved a memorability of 25% when two random characters were inserted. These findings further justify the need to evaluate new techniques (see Section 2.2) to improve users' memorability of system-generated data.

### 2.2 Memorability and Gamification

Bonneau and Schechter (2014) found that most users can memorize passwords when using tools that support learning over time. Atkinson and Shiffrin (1968) proposed a cognitive memory model, in which, new information is transferred to short-term memory through sensory organs. The short-term memory holds this new information as mental representations of selected parts of the information. This information is only passed from short-term to long-term memory when it can be encoded through cue-association (Atkinson and Shiffrin 1968) (e.g. when we see a cat it reminds us of our first cat). This encoding helps people remember and retrieve the stored information over a long period of time. These encodings are strengthened through constant rehearsals. Psychology research (Paivio et al. 1968; Thornton 2001) found that humans are better at remembering images than text (i.e. the picture superiority effect). The picture superiority effect has also been extensively used in usable security to research graphical authentication mechanisms (De Angeli et al. 2005; Denning et al. 2011; Stobert and Biddle 2013; Castellucia et al. 2017). Hence, in our research we use these psychology findings to design a game that nudges users' memorability of stronger answers to security questions.

Besides the previously described work, in our research we use a game-based learning approach because previous work in the security field (Arachchilage and Love 2013) has successfully used this approach to educate users about the susceptibility to phishing attacks, to teach users to be less prone to these types of security vulnerabilities (Arachchilage et al. 2012, 2014, 2016). Thus, our main contribution to the field of fall-back authentication is to investigate whether or not the proposed serious game has the potential of improving users' memorability of stronger answers to security questions.

## 3　Serious Game

This section describes the game design, security questions and system-generated profiles used in our user evaluation.

### 3.1　Game Design

Using the popular "*4 Pics 1 Word*" mobile game, we create encoding associations between cues and answers to security questions through the picture-based nature of this game and by adding verbal cues. The game functions similarly to the "*4 Pics 1 Word*" mobile game, meaning that the game asks players to pick the word that relates the given pictures (e.g. for the pictures in Figure 1a the relating word would be "Walk"). However, at certain intervals, the game asks players to solve challenges based on answers to their security questions. These challenges could either be based on users' own answers to security questions or on answers based on system-generated profiles. The game provides players with 12 letters to assist them with solving the challenge (as in the original "*4 Pics 1 Word*" mobile game). For each given answer, players are either rewarded or deducted points (10 for standard challenges, 15 for recognition security questions challenges and 20 for recall security questions challenges – these numbers were selected to motivate players to solve security question challenges). Points can be used to obtain hints to help solving more difficult challenges (deduction of 50 points for each hint, as in the original game).

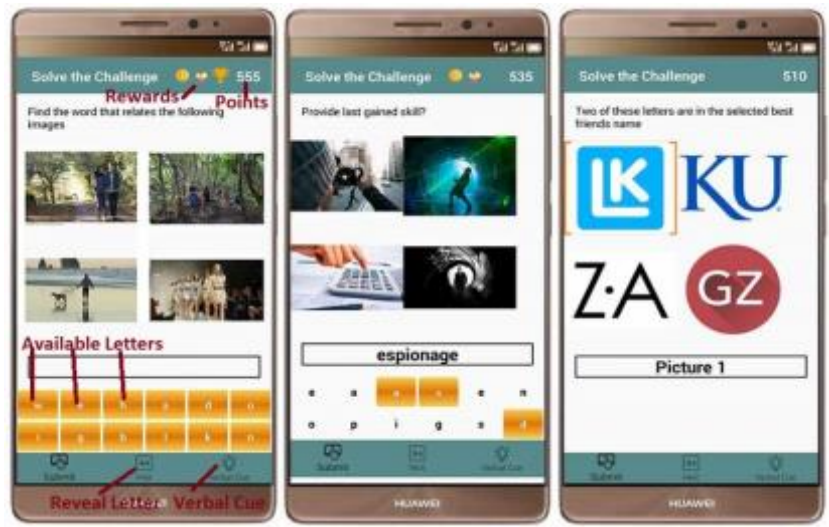

*Figure 1: Challenges:　a) standard, b) recall, c) recognition.*

The Generate-recognize theory (Anderson and Bower 1972) states that recognition (i.e. remembering contextual information when a focus is provided (Hollingworth 1913)) is easier and faster to perform than recall (i.e. remembering a specific focus when context is provided (Hollingworth 1913)). Thus, for security questions challenges, in the proposed game we use both recall and recognition challenges (see Figure 1b and 1c) because having only recognition challenges would have lowered the security level of the game, since the answer space would have been very limited. Therefore, the players are provided with more recognition challenges (i.e. select 1 of 4 pictures) at the beginning, with the purpose of encoding into long-term memory the associations between cues (pictures) and information (answers to security questions). Then, as the player gets used to the game and learns the answers the game starts showing more recall (i.e. provide exact answer to security questions) rather than recognition challenges.

Psychology research (Anderson and Bower 1972) has shown that it is difficult to remember information spontaneously without having any kind of memory cues. Hence, besides showing the 12 letters we added a feature that shows verbal cues about each picture. This feature can be enabled by using the points (i.e. when the game player earns 50 points throughout the game) that are gathered when solving other game challenges. This feature was added so that players could focus their attention on associating the words with the corresponding cues (pictures). We hypothesize that this feature should also help to process and encode the information in memory, to store it in long-term memory (Al-Ameen at al. 2015). To reduce the vulnerabilities towards potential guessing attacks, our serious game has the following features:　(1) does not show the length of the word that needs to be guessed; (2) does not show the correct answer when the wrong answer is provided; and (3) does not provide any hints for recognition challenges since the answer space is already very limited. These features make the game more difficult, but we argue that they minimize the security vulnerabilities of the game.

## 3.2 Security Questions

We used previous research (Rabkin 2008; Bonneau et al. 2010; Micallef and Just 2011; Bonneau et al. 2015) to define the security questions categories listed in Table 1. Names, favourites and places were selected because they are the most popular types of security questions found on websites (Rabkin 2008; Bonneau et al. 2010). Numbers were selected because they could potentially be the most secure security questions (Bonneau et al. 2015). Characteristics were added to make the system-generated data look similar to a profile (Micallef and Just 2011). To cover a wide range of questions we arbitrary selected 3 questions for each category (see Table 1). During the evaluation we asked participants to select 3 security questions because Renaud and Just (2010) found that posing 3 or more questions serially would be more secure, since it is difficult to guess all 3 answers irrespective of how close the attacker is to the victim. In our studies we do not use freely chosen security questions (e.g. the user defines his own answers to security questions) with free-form answers because Just and Aspinall (2010) reported serious concerns over the usability of these security questions (e.g. difficult to precisely remember the given answers).

| Type | Security Questions |
| --- | --- |
| Names | Mother's maiden name, Father's middle name, Best friends name. |
| Favourites | Favourite pet, Favourite food, Favourite hobby. |
| Numbers | Last 6 digits Visa no, Last 6 digits Phone number, Vehicle registration number. |
| Places | High school city name, College city name, First work city name. |
| Characteristics | First occupation, Last gained skill, Main Weakness. |

*Table 1. Security questions.*

## 3.3 System-generated Profiles

Figure 2 shows the system-generated profiles that we used in our user evaluation. We defined these system-generated profiles using Fake Name Generator (Corban Works 2006). We selected the attributes of these profiles by combining the attributes that were used by Micallef and Just (2011) to the list of security questions that we defined in Table 1. We defined a male and a female profile so that we cover the two most common genders. Since the design of a system-generated profile is not the main focus of this research, further research needs to be conducted to identify the optimal attributes that are required for a system-generated profile to be used to answer security questions.

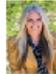

*Figure 2: System-generated profiles*

## 4　User Evaluation

To answer our research questions (**RQ1** and **RQ2**), we conducted a lab evaluation which consisted of two sessions (Session 1 – participants selected security questions and answers, Session 2 – participants memorized answers to security questions, played game and remembered answers to security questions). We used a between subjects design to compare and contrast our two groups of participants. Participants were randomly split into 2 equal groups. Group 1 participants were provided with a list of security questions. Then, they were asked to come up with the answers to those security questions by themselves. Group 2 participants were given 2 system-generated profiles (see Figure 2) and they were asked to choose one profile. Then, they were asked to choose security questions/answers from the chosen profile. Prior to starting the evaluation we obtained ethical approval from our University's Ethics Committee.

### 4.1　Procedure

Before starting the evaluation we conducted a small pilot study with 4 participants to evaluate the setup. Since the pilot study did not highlight any problems (we just removed 4 questions from pre-evaluation questionnaire) we included the 4 pilot participants in the main user evaluation. Session 1 was conducted separately because we needed time to configure the serious game based on the answers that were selected by the participants. In Session 1, participants were briefed on the evaluation, were asked to sign the consent form and selected security questions and answers. Participants also completed a pre-evaluation questionnaire, consisting of demographic information together with details about participants' experience with security questions and mobile phones. After configuring the game, this always happened on the same day as Session 1, we met with participants to conduct Session 2.

In Session 2, participants were asked to conduct the following three tasks: Task 1 - memorize answers to security questions; Task 2 - play a game; and Task 3 - remember answers to security questions. All participants memorized the answers to the security questions within 5 minutes. To understand the impact of using system-generated profiles on memorizing answers, playing the game and remembering answers (**RQ2**), participants completed the standard NASA task load index (NASA/TLX) questionnaire (Hart & Staveland 1988) after conducting every task (Tasks 1-3). We used the NASA/TLX metrics to evaluate workload because it is the standard metrics that is used in mobile HCI (Brewster 2002; Micallef et al. 2016) and usable security (Juang et al. 2012; Sherman et al. 2014) to evaluate mental demand, temporal demand, physical demand, performance, effort and frustration of using a system. After memorization, participants conducted an arithmetic distraction exercise (Bateman 2007a) for 5 minutes (Juang et al., 2012). At this stage participants were handed a mobile device (Samsung Galaxy S4) with the game described in Section 3 and configured with the answers to the security questions that participants selected in Session 1. We first went through a test game together with the participants (to show them how to play the game) and then we told them to play the game on their own.

The game started by picking a random standard challenge from a pool of 7 standard challenges (all participants experienced the same standard challenges but in a random order). After completing a standard challenge, the game player was given/deducted points. Afterwards, the challenge was removed from the pool of available challenges. At this stage the player was presented with a randomly selected recognition security questions challenge (based on the security answers that they selected prior to playing the game). The player continued to be presented with alternate standard and recognition challenges until they completed the 3 recognition security questions challenges. After that, the player was prompted with alternate standard and recall challenges until all 3 recall security questions challenges were completed. This is where the game ended. In total, each player completed 7 standard challenges, 3 recognition and 3 recall security questions challenges.

After playing the game, the participants completed the NASA/TLX to measure the workload of playing the game. After completing another 5 minutes distraction task (Bateman 2007b) participants were asked to write down the answers to their security questions to understand the short-term effect of using the game to remember the answers for both groups (**RQ1**).

### 4.2　Participants

We recruited 20 participants (5 females, 15 males) through word of mouth and personal connections. The mean age was 29 (22-45), med=28. 16 participants were post-graduate students and the rest (4) were employed full-time. Only 2/20 participants reported that they were not experienced and confident with using security questions on online websites. All participants (20/20) self-reported that they owned a smartphone for more than three years.

## 5 Results

In this section we present how the results of the user evaluation described in Section 4 answer our research questions (**RQ1** and **RQ2**). Since we used independent samples, we tested for statistical differences between our groups (Group 1 and Group 2) using T-Test (for independent samples) when data was normalized and Mann-Whitney U test when data was not normalized (< 0.05 using Shapiro-Wilk normality test). We assume that statistically significant differences are achieved when p < 0.05.

### 5.1 Does the game have the potential to help users remember answers to security questions?

After playing the game we asked all participants to write down the answers to the security questions that they selected in Session 1. We wanted to understand whether participants remember them (short-term) after conducting two arithmetic distraction tasks and playing a game in between. All participants (20/20) remembered the answers. All participants (10/10) that used their own answers to security questions (Group 1) reported that their answers were based on their own life experiences. Hence, although it was highly likely that these participants would remember (short-term) the answers to the security questions that are based on their own life experiences, it is still interesting to notice that no-one used fake answers. With regards to Group 2, it is interesting to note that 4/10 participants that answered security questions based on a system-generated profile failed to solve recall security questions game challenges. However, despite the game did not provide them with the correct answer (due to security reasons), they still provided the correct answer when they were asked to write down their answer. These findings indicate that the proposed serious game has the potentially of helping users which use stronger answers to security questions to remember their answers, even in the long-term.

### 5.2 What's the impact of system-generated profiles on memorizing answers, playing a game and remembering answers to security questions? (RQ2)

To understand the impact of using system-generated data for answers to security questions we measured the workload of memorizing answers, playing the game and remembering answers. Workload was collected using the standard NASA/TLX questionnaire (Hart & Staveland 1988), which measures mental demand, physical demand, temporal demand, effort, performance, and frustration on a scale of 0 to 100.

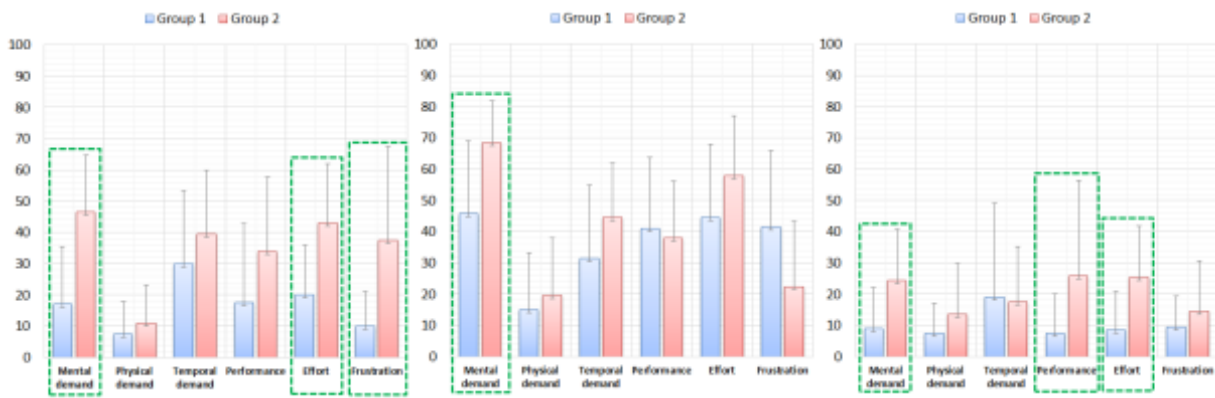

*Figure 3: Workload for a) Task 1 - Memorizing Answers, b) Task 2- Playing Game and c) Task 3 - Remembering Answers for Groups 1 and 2. Significant differences are shown in green dashed boxes.*

As shown in Figure 3a, participants evaluated the workload of memorizing answers to security questions (Task 1), which were based on their own security answers (Group 1) to be low in all measures. This result is related to the fact that all participants reported that they use answers based on their own life experiences to answer these questions. Although, for all measures participants that based their answers on a system-generated profile (Group 2) reported higher workload (see Figure 3a), only *mental demand* (t(18)=-3.594, p=0.002), *effort* (t(18)=-2.939, p=0.009) and *frustration* (t(18)=-2.726, p=0.014) were significantly higher than the results obtained for Group 1. This means that memorizing system-generated answers for security questions was considered to have a medium workload (closer to low) for *mental demand*, *effort* and *frustration* while the other measures were considered to be low and similar to using their own answers to security questions.

Mental demand after the game play activity (Task 2) was evaluated to be significantly higher by participants that used a system-generated profile to answer security questions (t(18)=-2.685, p=0.015). All the other measures of the game activity were evaluated similarly by both groups (no significant differences (see Figure 3b)). Except for *physical demand* and *frustration* which were evaluated to be low, the other measures (*mental demand*, *temporal demand*, *performance* and *effort*) were evaluated to have medium workload. This result might be related to the challenging aspect of the game, since on average participants failed to solve 2 standard challenges out of 7.

The task of remembering answers to security questions (Task 3) was evaluated to be low for all measures for both groups (see Figure 3c). However, participants that had to remember answers to security questions based on a system-generated profile (Group 2) found the *mental demand* (U=18, p=0.014), *performance* (U=24.5, p=0.048) and *effort* (U=14, p=0.006) to require significantly more workload than participants that remembered their own answers (Group 1). There were no significant differences between the groups for the other measures (*physical demand*, *temporal demand* and *frustration*). These findings imply that despite requiring more effort and mental demand, participants in Group 2 still required a low workload to remember answers to security questions using a system-generated profile and in the short-term there was not much difference compared to the other group.

We did not find any significant differences in game performance (similar amounts of solved challenges, used hints and time taken to play the game). These findings suggest that playing the game using challenges of system-generated answers to security questions did not require much extra effort because the workload and game performance were mostly similar to using answers to security questions based on participants own lives. Also, although using system-generated data significantly impacted the workload of memorizing and remembering answers, the workload was still evaluated to be medium for memorizing answers (which is a one-time task) and low for remembering answers.

# 6　Discussion

In this section, we discuss how our results could help bridge the trade-off between security and usability (in terms of memorability (Just and Aspinall 2009)) for security questions.

## 6.1　Improving memorability through serious games

Previous work has implemented novel graphical authentication schemes (Denning et al. 2011; Stobert and Biddle 2013; Castellucia et al. 2017) or used mnemonics (Juang et al. 2012) and games to improve the memorability of passwords (Tao and Adams 2008; Malempati and Mogalla 2011; McLennan et al. 2017). Since the system-generated data that we used in our studies (see Figure 2) is completely different from the passwords/schemes used by previous work, we cannot conduct any direct memorability comparisons. However, since the proposed serious game seems to have helped participants who were using system-generated profiles to remember their answers, even when they failed to solve some of the recall challenges presented by the game, we argue that from a high-level perspective our memorability results seem to confirm the effectiveness of using serious games to improve memorability of answers to security questions. More specifically, our findings seem to indicate that the proposed serious game could potentially lead to improve the long-term memorability of answers to security questions (this still has to be evaluated in a longitudinal field study). This improvement in memorability would directly improve the usability of stronger answers to security questions (Just and Aspinall 2009). Hence, this potential improvement in memorability of answers to security questions shows that the proposed serious game (which was designed using memorability concepts (Atkinson and Shiffrin 1968)) could eventually help reduce the trade-off between usability and security in fall-back authentication.

## 6.2　Impact of using system-generated profiles

When comparing the workload results to other work, we found that the workload (e.g. mental demand, temporal demand, physical demand, effort, etc.) of memorizing answers to security questions based on system-generated profiles is slightly higher than when using user-generated free-form gestures (gestures that allow all fingers to draw a path on an empty screen with requiring a grid) for authentication (Sherman et al. 2014). Remembering system-generated answers is comparable to using free-form gestures. When comparing our workload results to a study which assessed password reset policies at a university (Parkin et al. 2015), we found that using system-generated answers requires less workload than registering for a password recovery policy and when authenticating to recover passwords. Hence, our work also reveals that using system-generated profiles to answer security questions did not have any significant short-term effect on any major aspects of our evaluation or compared to other work in the area. Therefore, we argue that the use of system-generated data to answers security questions could play an important role in reducing the vulnerabilities (to social engineering attacks) of our online accounts (Shay et al. 2012) without requiring much extra effort (in the-short term), when compared to using our own answers to security questions.

## 7　Limitations and Future Work

One of the main limitations of our evaluation is that we only evaluated the short-term memorability of the game to nudge users' memorability of stronger answers to security questions. Hence, future studies need to confirm whether or not our findings could be extended to long-term memorability. Another limitation is that we primarily focused on the 18-35 age group, since in Australia this is the demographic that uses mobile phones the most (95%) (Poushter 2016). Although, other research (McLennan et al. 2017) found that serious games in security education are considered to be fun by different populations, we still plan to investigate different age groups to determine whether our findings also extend to different populations. We will also conduct a security evaluation of the proposed serious game to determine the security vulnerabilities that need to be addressed to achieve the required security level. Afterwards, we will conduct a longitudinal field study to determine how much training is required, so that users learn the answers to their security questions, when using the proposed serious game.

## 8　Conclusions

The main outcome of this research is that our participants remembered (short-term) their answers to security questions (**RQ1**) after the game play activity (including participants that used system-generated profiles, see Figure 2). Our work also revealed that using system-generated profiles to answer security questions did not have any significant short-term effect on playing the game and remembering answers to security questions, compared to users that used their own answers (**RQ2**). Thus, the main contribution of our work is that our findings indicate that the proposed serious game could potentially lead to improve the users' long-term memorability of stronger answers to security questions. This improvement in memorability would directly improve the usability of using strong answers to security questions (Just and Aspinall 2009). Hence, we strongly believe that the potential improvement in memorability achieved through the use of a serious game (which uses memorability concepts) could eventually help reduce the trade-off between usability and security in fall-back authentication.

## 9　References